\documentclass[a4paper,11pt]{article}

\usepackage{jheppub_mod} 
\usepackage[T1]{fontenc} 

\usepackage{amsmath,booktabs}
\usepackage{graphicx}
\usepackage{xspace}
\usepackage{color}
\pagecolor{white}
\usepackage[dvipsnames]{xcolor}
\usepackage{url}
\usepackage[utf8]{inputenc}
\usepackage{epstopdf}
\usepackage{hyperref}
\usepackage{multirow}
\usepackage{gensymb}

\newcommand{\mpi}{M_\pi}
\newcommand{\mpii}{M_{\pi^0}}
\newcommand{\beq}{\begin{equation}}
\newcommand{\eeq}{\end{equation}}
\newcommand{\diff}{\text{d}}
\newcommand{\eps}{\epsilon}

\newcommand{\mw}{M_\omega}
\newcommand{\Gw}{\Gamma_\omega}

\newcommand{\mphi}{M_\phi}
\newcommand{\Gphi}{\Gamma_\phi}

\newcommand{\epsrw}{\eps_{\omega}}
\newcommand{\gwg}{g_{\omega\gamma}}

\newcommand{\GeV}{\,\text{GeV}}
\newcommand{\MeV}{\,\text{MeV}}

\renewcommand{\Im}{\text{Im}\,}
\renewcommand{\Re}{\text{Re}\,}

\newcommand{\F}{\mathcal{F}}
\newcommand{\A}{\mathcal{A}}

\allowdisplaybreaks[1]

\title{Extracting the chiral anomaly from $\boldsymbol{e^+e^-\to 3\pi}$}

\author[a]{Martin Hoferichter,}
\author[b]{Bai-Long Hoid,}
\author[c]{and Bastian Kubis}

\affiliation[a]{Albert Einstein Center for Fundamental Physics, Institute for Theoretical Physics, University of Bern, Sidlerstrasse 5, 3012 Bern, Switzerland}

\affiliation[b]{Institut f\"ur Kernphysik and PRISMA$^+$  Cluster of Excellence, Johannes Gutenberg Universit\"at,  55099 Mainz, Germany}

\affiliation[c]{Helmholtz-Institut f\"ur Strahlen- und Kernphysik (Theorie) and \\
Bethe Center for Theoretical Physics, Universit\"at Bonn, 53115 Bonn, Germany}

\emailAdd{hoferichter@itp.unibe.ch}
\emailAdd{lonbai@uni-mainz.de}
\emailAdd{kubis@hiskp.uni-bonn.de}

\abstract{The strength of the interaction of three pions and a photon, $F_{3\pi}$ is predicted by the axial anomaly in terms of the pion decay constant, a relation that is frequently used to constrain low-energy radiative processes involving pions, but only tested experimentally  at the $10\%$ level. Here, we present a new avenue to test this prediction, via a fit of a dispersive description of the $\gamma^*\to3\pi$ amplitude to data for $e^+e^-\to 3\pi$. From the global fit to SND, CMD-2, and BaBar data we obtain $F_{3\pi}=33.1(1.7)\GeV^{-3}$, in agreement with the chiral prediction at the level of $5\%$. We also consider dispersive fits to the recent data by Belle~II, in which case we observe tensions with the dispersive constraints, the width parameters of $\omega$ and $\phi$, and the chiral anomaly.}

\begin{document}
\maketitle
	
\section{Introduction}
\label{sec:intro}

The interaction strength of pions with photons in odd intrinsic-parity processes at low energies is predicted
 by the Wess--Zumino--Witten (WZW) anomaly~\cite{Wess:1971yu,Witten:1983tw,Adler:1971nq,Terentev:1971cso,Aviv:1971hq}, most prominently, for the $\pi^0\to\gamma\gamma$ decay the amplitude in the chiral limit is determined by the pion decay constant $F_\pi=92.32(10)\MeV$~\cite{ParticleDataGroup:2024cfk} according to
\beq
\label{pi0gg}
F_{\pi\gamma\gamma}^\text{WZW}\equiv\frac{1}{4\pi^2F_\pi}=0.2744(3)\GeV^{-1},
\eeq
where the dominant quark-mass corrections have been absorbed into the physical value of $F_\pi$. Accordingly, already this leading-order expression agrees perfectly with the PrimEx-II measurement $F_{\pi\gamma\gamma}^\text{exp}=0.2754(21)\GeV^{-1}$~\cite{PrimEx-II:2020jwd}, in such a way that higher-order corrections~\cite{Bijnens:1988kx,Goity:2002nn,Ananthanarayan:2002kj,Kampf:2009tk,Gerardin:2019vio} create a mismatch to next-to-next-to-leading-order chiral perturbation theory $F_{\pi\gamma\gamma}^\text{NNLO}=0.2801(17)\GeV^{-1}$\cite{Kampf:2009tk,Gerardin:2019vio}  at the level of $1.7\sigma$.\footnote{The lattice-QCD calculation of Ref.~\cite{Gerardin:2019vio} supports the approximation for the anomalous chiral low-energy constants of next-to-leading order~\cite{Bijnens:2001bb},  $C_7^W\ll C_8^W$ as assumed in Ref.~\cite{Kampf:2009tk} ($C_8^W$ then inferred from $\eta\to \gamma\gamma$ via $SU(3)$), disfavoring the opposite conclusion $C_7^W> C_8^W$ in Ref.~\cite{GrillidiCortona:2015jxo}. In the $U(3)$ analysis of Refs.~\cite{Gao:2022xqz,Gao:2024vkw}, the tension in $F_{\pi\gamma\gamma}$ reduces, but a global fit to $F_{P\gamma\gamma}$, $P=\pi,\eta,\eta'$, is used to determine two unknown low-energy constants.} Taking the difference between $F_{\pi\gamma\gamma}^\text{exp}$ and $F_{\pi\gamma\gamma}^\text{NNLO}$, the WZW prediction would thus be validated at a level of at least $2\%$, or $0.8\%$ when considering the experimental precision alone.

The main quantity of interest in this work is the amplitude for the next-simplest process, $\gamma\to3\pi$~\cite{Hoid:2025CD,Moinester:2024lwl}.
In analogy to Eq.~\eqref{pi0gg}, we again absorb the dominant quark-mass corrections to the pion-decay constant into the chiral prediction
\beq
\label{F3pi_WZW}
F_{3\pi}^\text{WZW}\equiv \frac{1}{4\pi^2 F_\pi^3}=32.19(10)\GeV^{-3}.
\eeq
Also in this case, higher-order corrections have been studied in considerable detail~\cite{Bijnens:1989ff,Alkofer:1995jx,Holstein:1995qj,Hannah:2001ee,Truong:2001en,Ametller:2001yk,Benic:2011rk,Bijnens:2012hf}, the most important of which concern an additional quark-mass renormalization beyond $F_\pi$ that changes the physically accessible constant to~\cite{Bijnens:1989ff}
\beq
\label{C_A_chiral}
C_A=F_{3\pi}\times 1.066(10),
\eeq
where the uncertainty is derived from variation of the scale in the chiral logarithm between $[M_\rho/2,2M_\rho]$ and the amount of universality violation in the $\rho\pi\pi$ and $\rho\gamma$ couplings~\cite{Hoferichter:2012pm,Hoferichter:2017ftn}. At present, experimental determinations from the low-energy tail of the Primakoff reaction, $F_{3\pi}=35.3(4.0)\GeV^{-3}$~\cite{Antipov:1986tp} (including radiative corrections from Ref.~\cite{Ametller:2001yk}), and $\pi^-e^-\to \pi^-e^-\pi^0$, $F_{3\pi}= 31.7(3.6)\GeV^{-3}$~\cite{Giller:2005uy}, agree with the chiral prediction~\eqref{F3pi_WZW} at the level of $10\%$. To improve,  it was suggested in Refs.~\cite{Hoferichter:2012pm,Hoferichter:2017ftn} to employ a dispersive representation to leverage Primakoff data including the $\rho(770)$ peak for the anomaly extraction, and preliminary results at COMPASS, $F_{3\pi}= 34.0(2.0)\GeV^{-3}$~\cite{Ecker:2023qae,Maltsev:2025CD}, suggest a precision around $6\%$, to be updated soon with the final COMPASS result. The process $\gamma^{(*)}\pi\to\pi\pi$ has also triggered interest in lattice QCD~\cite{Briceno:2015dca,Briceno:2016kkp,Alexandrou:2018jbt}, as the simplest scattering process involving an external current and a hadronic resonance. To extract the chiral anomaly from these calculations, a substantial chiral extrapolation is required, leading to $F_{3\pi}=38(19)\GeV^{-3}$~\cite{Niehus:2021iin}, consistent with Eq.~\eqref{F3pi_WZW}, but only within a $50\%$ uncertainty. Finally, $F_{3\pi}$ has been studied recently using Dyson--Schwinger equations~\cite{Xing:2024bpj,Miramontes:2025ofw}.

Meanwhile, $F_{3\pi}^\text{WZW}$ has been used extensively in recent years in the context of hadronic corrections to the anomalous magnetic moment of the muon $a_\mu$~\cite{Aoyama:2020ynm,Aoyama:2012wk,Aoyama:2019ryr,Czarnecki:2002nt,Gnendiger:2013pva,Davier:2017zfy,Keshavarzi:2018mgv,Colangelo:2018mtw,Hoferichter:2019mqg,Davier:2019can,Keshavarzi:2019abf,Hoid:2020xjs,Kurz:2014wya,Melnikov:2003xd,Colangelo:2014dfa,Colangelo:2014pva,Colangelo:2015ama,Masjuan:2017tvw,Colangelo:2017qdm,Colangelo:2017fiz,Hoferichter:2018dmo,Hoferichter:2018kwz,Gerardin:2019vio,Bijnens:2019ghy,Colangelo:2019lpu,Colangelo:2019uex,Blum:2019ugy,Colangelo:2014qya,Aliberti:2025beg}, entering both dispersive calculations of the pion-pole contribution to hadronic light-by-light scattering~\cite{Schneider:2012ez,Hoferichter:2012pm,Hoferichter:2014vra,Hoferichter:2018dmo,Hoferichter:2018kwz,Hoferichter:2021lct,Ludtke:2024ase,Hoferichter:2024vbu,Hoferichter:2024bae,Hoferichter:2025yih,Hoferichter:2025fea} and the cross sections for $e^+e^-\to3\pi,\pi^0\gamma$~\cite{Hoferichter:2019mqg,Hoid:2020xjs,Hoferichter:2023bjm,Hoferichter:2023sli}, and similar anomaly constraints also pertain to $\eta$, $\eta'$~\cite{Stollenwerk:2011zz,Hanhart:2013vba,Kubis:2015sga,Holz:2015tcg,Holz:2022hwz,Holz:2022smu,Holz:2024lom,Holz:2024diw} and related photon--kaon processes~\cite{Vysotsky:2015mks,Dax:2020dzg,Stamen:2024ocm}. Accordingly, an improved determination of $F_{3\pi}$ would allow an important consistency check and foster confidence in the robustness of the imposed chiral constraints. In this paper, we therefore study the opposite strategy, and try and \emph{extract} $F_{3\pi}$ from $e^+e^-\to 3\pi$ data instead of using it as a constraint.\footnote{We thank Andrei Rabusov for this suggestion.}

To this end, we first summarize the dispersive formalism from Refs.~\cite{Hoferichter:2019mqg,Hoferichter:2023bjm} in Sec.~\ref{sec:formalism}, focusing on the role of the chiral anomaly $C_A$ in constraining the subtraction function $a(q^2)$. In Sec.~\ref{sec:anomaly}, we perform a global fit of $C_A$ to data from SND~\cite{Achasov:2000am,Achasov:2002ud,Achasov:2003ir,SND:2020ajg}, CMD-2~\cite{Akhmetshin:1995vz,Akhmetshin:1998se,Akhmetshin:2003zn,Akhmetshin:2006sc}, and BaBar~\cite{BABAR:2021cde}. In Sec.~\ref{sec:BelleII}, we extend these fits to the recent data by Belle II~\cite{Belle-II:2024msd}, both with and without $C_A$ as an additional fit parameter. We conclude in Sec.~\ref{sec:conclusions}.

\section{Dispersive formalism}
\label{sec:formalism}

To describe the cross section for $e^+e^-\to3\pi$ we use the dispersive formalism developed in Refs.~\cite{Hoferichter:2014vra,Hoferichter:2019mqg,Hoferichter:2023bjm}. First, the matrix element for $\gamma^*(q)\rightarrow \pi^+(p_+)\pi^-(p_-)\pi^0(p_0)$  is decomposed into
\begin{equation}
    \langle0|j_\mu(0)|\pi^+(p_+)\pi^-(p_-)\pi^0(p_0)\rangle = -\epsilon_{\mu\nu\alpha\beta}p_+^\nu p_-^\alpha p_0^\beta \F(s,t,u;q^2),
\end{equation}
where $q=p_++p_-+p_0$,
\beq
s=(p_++p_-)^2,\qquad t=(p_-+p_0)^2,\qquad u=(p_++p_0)^2,
\eeq
$s+t+u=3M_\pi^2+q^2$, and the scalar function $\F$ is inferred from the reconstruction theorem
\begin{equation}
\label{reconstruction}
    \F(s,t,u;q^2)
= \F(s,q^2)+\F(t,q^2)+\F(u,q^2)
\end{equation}
in combination with the partial-wave expansion~\cite{Jacob:1959at}
\beq
\F(s,t,u;q^2)=\sum \limits_{\ell \text{\,odd}} f_\ell(s,q^2) P'_\ell(z_s),
\eeq
where $P_\ell'(z)$ are the derivatives of the Legendre polynomials.
Moreover, Eq.~\eqref{reconstruction} applies when the discontinuities of $F$- and higher partial waves can be neglected, which is the case below the $\rho_3(1690)$ resonance~\cite{Niecknig:2012sj,Hoferichter:2017ftn,Hoferichter:2019mqg}.
Using the kinematic variables
\begin{align}
        z_s&=\cos\theta_s =\frac{t-u}{\kappa(s,q^2)}, \qquad
        \kappa(s,q^2)=\sigma_\pi(s)\lambda^{1/2}(q^2,M_\pi^2,s),\notag\\
     \lambda(x,y,z)&=x^2+y^2+z^2-2(xy+yz+xz),\qquad  \sigma_\pi(s)=\sqrt{1-\frac{4\mpi^2}{s}},
\end{align}
the $e^+e^-\to 3\pi$ cross section becomes
\beq
\label{eq:epemcross1}
\sigma_{e^+ e^- \to 3\pi}(q^2) = \alpha^2\int_{s_\text{min}}^{s_\text{max}} \diff s \int_{t_\text{min}}^{t_\text{max}} \diff t \,
\frac{s[\kappa(s,q^2)]^2(1-z_s^2)}{768 \, \pi \, q^6}  \, |\F(s,t,u;q^2)|^2,
\eeq
integrated within the boundaries
\begin{align}
s_\text{min} &= 4 M_\pi^2, \qquad\qquad \,s_\text{max} = \big(\sqrt{q^2}-M_\pi \big)^2, \notag \\
t_\text{min/max}&= (E_-^*+E_0^*)^2-\bigg( \sqrt{E_-^{*2}-M_\pi^2} \pm  \sqrt{E_0^{*2}-M_\pi^2} \bigg)^2,\notag\\
E_-^*&=\frac{\sqrt{s}}{2},\qquad E_0^*=\frac{q^2-s-M_\pi^2}{2\sqrt{s}}.
\end{align}
The partial waves are calculated in the Khuri--Treiman formalism~\cite{Khuri:1960zz}, which allows one to include $\pi\pi$ rescattering corrections, while the global normalization $a(q^2)$ is not predicted. We employ the parameterization
\beq
\label{eq:a-par}
a(q^2)=\alpha_A+\frac{q^2}{\pi}\int_{s_\text{thr}}^{\infty} \diff s'\frac{\Im\A (s')}{s'(s'-q^2)}+C_p(q^2),
\eeq
where the first term
\beq
\alpha_A=\frac{C_A}{3}
\eeq
entails the information about the WZW anomaly. The other terms include explicit resonance contributions and a conformal polynomial $C_p(q^2)$, see Refs.~\cite{Hoferichter:2019mqg,Hoferichter:2023bjm} for explicit expressions, which generate imaginary parts above thresholds $s_\text{thr}$ and $s_\text{inel}$, respectively. In practice, they are chosen as $s_\text{thr}=\mpii^2$ (due to the $\omega\to\pi^0\gamma$ decay) and $s_\text{inel}=1\GeV^2$ (motivated by the nearby $\bar K K$ threshold).
Moreover, the resonances are described via dispersively improved Breit--Wigner-type parameterizations~\cite{Lomon:2012pn,Moussallam:2013una,Zanke:2021wiq,Crivellin:2022gfu} to ensure that the correct analytic structure is maintained.

\begin{table}[t!]
	\centering
	\footnotesize
	\renewcommand{\arraystretch}{1.3}
	\begin{tabular}{lcccccc}
	\toprule
	& \multicolumn{3}{c}{$\delta_\eps=0\degree$} & \multicolumn{3}{c}{$\delta_\eps=3.5\degree$}\\
	$p_\text{conf}$ & $2$ & $3$ & $4$ & $2$ & $3$ & $4$\\
	$\chi^2/\text{dof}$ & $504.9/368$ & $494.7/367$ & $481.9/366$ & $512.7/368$ & $497.2/367$ & $479.4/366$\\
	& $=1.37$ & $=1.35$ & $=1.32$ & $=1.39$ & $=1.35$ & $=1.31$\\
	$p$-value & $3\times 10^{-6}$ & $1\times 10^{-5}$ & $4\times 10^{-5}$ & $8\times 10^{-7}$ & $7\times 10^{-6}$ & $6\times 10^{-5}$\\
	$\mw \ [\text{MeV}]$ & $782.70(3)$ & $782.70(3)$ & $782.70(3)$ & $782.69(2)$ & $782.70(3)$ & $782.70(2)$\\
	$\Gw \ [\text{MeV}]$ & $8.70(2)$ & $8.71(2)$ & $8.72(2)$  & $8.69(2)$ & $8.70(2)$ & $8.72(2)$\\
	$\mphi \ [\text{MeV}]$& $1019.21(1)$ & $1019.21(1)$ & $1019.21(1)$ & $1019.22(1)$ & $1019.21(1)$ & $1019.21(1)$\\
	$\Gphi \ [\text{MeV}]$ & $4.27(1)$ & $4.27(1)$ & $4.27(1)$  & $4.27(1)$ & $4.27(1)$ & $4.27(1)$\\
	$M_{\omega'} \ [\text{GeV}]$ & $1.445(10)$ & $1.436(23)$& $1.418(11)$& $1.444(9)$ & $1.432(5)$& $1.412(10)$\\
	$c_\omega \ [\text{GeV}^{-1}]$ & $2.93(1)$ & $2.93(1)$ & $2.96(2)$ & $2.93(1)$ & $2.94(2)$ & $2.97(2)$\\
	$c_\phi \ [\text{GeV}^{-1}]$ & $-0.380(1)$ & $-0.380(1)$ & $-0.381(2)$ & $-0.380(1)$ & $-0.380(1)$ & $-0.382(1)$\\
	$c_{\omega'} \ [\text{GeV}^{-1}]$ & $-0.24(3)$ & $-0.15(4)$ & $-0.23(5)$  & $-0.25(3)$ & $-0.13(4)$ & $-0.23(5)$\\
	$c_{\omega''} \ [\text{GeV}^{-1}]$ & $-1.77(4)$ & $-1.67(5)$ & $-1.59(6)$ & $-1.79(4)$ & $-1.68(5)$ & $-1.59(6)$\\
	$c_1 \ [\text{GeV}^{-3}]$ & $-0.18(5)$ & $-0.13(6)$ & $0.03(8)$ & $-0.15(5)$ & $-0.10(5)$ & $0.11(8)$\\
	$c_2 \ [\text{GeV}^{-3}]$ & $-1.21(3)$ & $-1.29(4)$ & $-1.43(6)$ & $-1.20(3)$ & $-1.30(4)$ & $-1.48(5)$\\
	$c_3 \ [\text{GeV}^{-3}]$ & --- & $-0.65(5)$ & $-0.57(6)$ & --- & $-0.65(4)$ & $-0.56(5)$\\
	$c_4 \ [\text{GeV}^{-3}]$ & --- & --- & $1.35(5)$ & --- & --- & $1.40(5)$ \\
	$10^4\times \xi_\text{CMD-2}$ & $1.4(5)$ & $1.3(5)$ & $1.3(5)$ & $1.4(5)$ & $1.3(5)$ & $1.3(5)$ \\
	$10^3\times \xi_\text{BaBar}$ & $1.3(2)$ & $1.3(2)$ & $1.3(2)$ & $1.3(2)$ & $1.3(2)$ & $1.3(2)$ \\
	$10^3\times \xi_\text{BaBar}' \ [\text{GeV}^{-1}]$ & $-2.3(3)$ & $-2.3(4)$ & $-2.3(3)$ & $-2.3(3)$ & $-2.3(4)$ & $-2.3(3)$ \\
	$10^3\times \Re\epsrw$ & $1.51(18)$ & $1.49(18)$ & $1.60(17)$ & $1.45(18)$ & $1.49(18)$ & $1.68(17)$\\
	$10^{10}\times a_\mu^{3\pi}|_{\leq 1.8\GeV}$ & $45.74(31)$ & $45.91(32)$ & $46.26(33)$ & $45.61(30)$ & $45.81(31)$ & $46.22(32)$\\
	$10^{10}\times a_\mu^\text{FSR}[3\pi]$ & $0.51(0)$ & $0.51(0)$ & $0.52(0)$ & $0.51(0)$ & $0.51(0)$ & $0.52(0)$\\
	$10^{10}\times a_\mu^{\rho\text{--}\omega}[3\pi]$ & $-2.70(31)$ & $-2.68(31)$ & $-2.91(30)$ & $-2.93(36)$ & $-3.04(37)$ & $-3.47(34)$\\
	\bottomrule
	\renewcommand{\arraystretch}{1.0}
	\end{tabular}
	\caption{Global fit to SND~\cite{Achasov:2000am,Achasov:2002ud,Achasov:2003ir,SND:2020ajg}, CMD-2$'$~\cite{Akhmetshin:1995vz,Akhmetshin:1998se,Akhmetshin:2003zn,Akhmetshin:2006sc}, and BaBar~\cite{BABAR:2021cde}, reproduced from Ref.~\cite{Hoferichter:2023bjm} (in the same conventions as given there). In particular, scale factors are not yet included in the uncertainties. $\mw$, $\Gw$, $\mphi$, $\Gphi$, $M_{\omega'}$ denote VP-subtracted vector-meson parameters, $c_\omega$, $c_\phi$, $c_{\omega'}$, $c_{\omega''}$ their residues (in units of $1/e=1/\sqrt{4\pi\alpha}$), $c_{1\text{--}4}$ parameters in the conformal polynomial, $\xi$ and $\xi'$ energy rescaling factors, and $a_\mu^{3\pi}$, $a_\mu^\text{FSR}[3\pi]$, $a_\mu^{\rho\text{--}\omega}[3\pi]$ give the total, FSR, and $\rho$--$\omega$ contribution to $a_\mu$ up to $1.8\GeV$, respectively.}
	\label{tab:fits_global}
\end{table}

Throughout, we use the extended formalism from Ref.~\cite{Hoferichter:2023bjm}, in which also $\rho$--$\omega$ mixing effects are taken into account, by means of the correction factor
\beq
\label{3pi_rw}
g_\pi(q^2)=1-\frac{\gwg^2\epsrw}{e^2}\Pi_\pi(q^2),
\eeq
to be multiplied to the $\omega$ contribution in $a(q^2)$,
where $|\gwg|=16.2(8)$, $e^2=4\pi\alpha$, and $\Pi_\pi(q^2)$ is the vacuum-polarization (VP) function for $\pi^+\pi^-$ intermediate states. This correction is derived in Ref.~\cite{Holz:2022hwz} using a coupled-channel formalism~\cite{Hanhart:2012wi,Ropertz:2018stk,VonDetten:2021rax,Heuser:2024biq}, which ensures that the resulting $\rho$--$\omega$ mixing parameter $\epsrw$ is consistent with the standard definition in $e^+e^-\to\pi^+\pi^-$. As shown in Ref.~\cite{Colangelo:2022prz}, this parameter can acquire a small phase due to radiative transitions such as $\rho\to\pi^0\gamma\to\omega$, but while the $\rho$--$\omega$ interference effect as such can be resolved in $e^+e^-\to 3\pi$, the sensitivity does not suffice to extract its phase. As in Ref.~\cite{Hoferichter:2023bjm}, we will therefore consider the cases $\delta_\eps=0\degree$ and $\delta_\eps=3.5\degree$, the latter motivated by narrow-resonance arguments~\cite{Colangelo:2022prz}, to gauge the sensitivity of the line shape to this phase. Finally, we also include the final-state-radiation (FSR) corrections in the data analysis, in the same form as derived in Ref.~\cite{Hoferichter:2023bjm}.

The sensitivity to the chiral anomaly now arises via the subtraction constant $\alpha_A$ in Eq.~\eqref{eq:a-par}, essentially, by leaving $C_A$ as free fit parameter instead of constraining it to the chiral prediction~\eqref{C_A_chiral}. Importantly, by the nature of the dispersive constraints, the sensitivity to $C_A$ is not limited to the low-energy region, since the sum rule
\beq
\label{eq:a_SR}
\alpha_A=\frac{1}{\pi}\int_{s_\text{thr}}^{\infty} \diff s'\frac{\Im{a}(s')}{s'}
=\frac{1}{\pi}\int_{s_\text{thr}}^{\infty} \diff s'\frac{\Im\A(s')}{s'}+\frac{1}{\pi}\int_{s_\text{inel}}^{\infty} \diff s'\frac{\Im{C_p}(s')}{s'}
\eeq
is imposed at each step in the fit~\cite{Hoferichter:2019mqg}.\footnote{In Ref.~\cite{Hoferichter:2014vra} we studied a variant in which the sum rule~\eqref{eq:a_SR} was not yet imposed, and indeed in this case the sensitivity to $C_A$ proved much weaker.}  Accordingly, $C_A$ emerges as a global property of the low-energy $e^+e^-\to3\pi$ cross section, which, apart from studying to which precision $C_A$ can be extracted, also renders this extended fit an interesting consistency check on $e^+e^-\to 3\pi$ data sets.

\begin{table}[t]
	\centering
	\footnotesize
	\renewcommand{\arraystretch}{1.3}
	\begin{tabular}{lccc}
	\toprule
	& Global fit & Belle II & Belle II (w/o reweighting)\\\midrule
	$\mw \ [\text{MeV}]$ & $782.697(32)(4)(4)[32]$ & $782.72(5)(0)(0)[5]$& $782.75(5)(0)(0)[5]$\\
	$\Gw \ [\text{MeV}]$ & $8.711(21)(12)(10)[26]$ & $8.43(8)(1)(2)[8]$ & $8.56(8)(1)(0)[8]$\\
	$\mphi \ [\text{MeV}]$&$1019.211(17)(4)(1)[17]$ & $1019.24(5)(1)(2)[5]$ & $1019.39(5)(1)(0)[5]$\\
	$\Gphi \ [\text{MeV}]$ & $4.270(13)(3)(1)[13]$ & $3.98(11)(2)(0)[11]$ &$4.44(11)(1)(0)[11]$\\
	$M_{\omega'} \ [\text{MeV}]$&$1436(26)(17)(6)[32]$&$1462(29)(5)(6)[30]$ & $1467(25)(3)(5)[26]$\\
	$\Re\epsrw\times 10^3$& $1.49(21)(11)(8)[25]$& $1.50(58)(7)(4)[59]$& $2.07(57)(7)(3)[57]$\\\midrule
	$10^{10}\times a_\mu^{3\pi}|_{\leq 1.8\GeV}$ & $45.91(37)(35)(13)[53]$ & $48.7(1.4)(0.1)(0.1)[1.4]$ & $48.5(1.4)(0.1)(0.2)[1.4]$\\
	$10^{10}\times a_\mu^\text{FSR}[3\pi]$ & $0.509(4)(6)(6)[9]$ & $0.541(15)(1)(5)[16]$ & $0.550(15)(1)(5)[16]$\\
	$10^{10}\times a_\mu^{\rho\text{--}\omega}[3\pi]$ & $-2.68(36)(22)(56)[70]$ & $-2.9(1.1)(0.1)(0.5)[1.2]$ & $-3.9(1.0)(0.1)(0.6)[1.2]$\\
	\bottomrule
	\renewcommand{\arraystretch}{1.0}
	\end{tabular}
	\caption{Fit parameters and derived quantities. The errors refer to statistics (including scale factor), truncation of the conformal polynomial, dependence on $\delta_\eps$, and quadratic sum, respectively.}
	\label{tab:derived_quantities}
\end{table}

\begin{table}[t]
	\centering
	\footnotesize
	\renewcommand{\arraystretch}{1.3}
	\begin{tabular}{lcccccc}
	\toprule
	& \multicolumn{3}{c}{$\delta_\eps=0\degree$} & \multicolumn{3}{c}{$\delta_\eps=3.5\degree$}\\
	$p_\text{conf}$ & $2$ & $3$ & $4$ & $2$ & $3$ & $4$\\
	$\chi^2/\text{dof}$ & $491.7/367$ & $490.1/366$ & $480.8/365$ & $489.4/367$ & $487.7/366$ & $479.3/365$\\
	& $=1.34$ & $=1.34$ & $=1.32$ & $=1.33$ & $=1.33$ & $=1.31$\\
	$p$-value & $1\times 10^{-5}$ & $1\times 10^{-5}$ & $4\times 10^{-5}$ & $2\times 10^{-5}$ & $2\times 10^{-5}$ & $5\times 10^{-5}$\\
	$\mw \ [\text{MeV}]$ & $782.70(3)$ & $782.70(1)$ & $782.70(1)$ & $782.70(1)$ & $782.70(3)$ & $782.70(2)$\\
	$\Gw \ [\text{MeV}]$ & $8.72(2)$ & $8.72(1)$ & $8.72(2)$  & $8.72(2)$ & $8.72(2)$ & $8.72(2)$\\
	$\mphi \ [\text{MeV}]$& $1019.21(1)$ & $1019.21(1)$ & $1019.21(1)$ & $1019.21(1)$ & $1019.21(1)$ & $1019.21(1)$\\
	$\Gphi \ [\text{MeV}]$ & $4.27(1)$ & $4.27(1)$ & $4.27(1)$  & $4.27(1)$ & $4.27(1)$ & $4.27(1)$\\
	$M_{\omega'} \ [\text{GeV}]$ & $1.444(10)$ & $1.440(13)$& $1.410(11)$& $1.444(10)$ & $1.440(10)$& $1.411(12)$\\
	$c_\omega \ [\text{GeV}^{-1}]$ & $2.94(2)$ & $2.94(1)$ & $2.96(1)$ & $2.96(1)$ & $2.96(2)$ & $2.97(2)$\\
	$c_\phi \ [\text{GeV}^{-1}]$ & $-0.381(1)$ & $-0.381(1)$ & $-0.381(1)$ & $-0.381(1)$ & $-0.381(2)$ & $-0.381(1)$\\
	$c_{\omega'} \ [\text{GeV}^{-1}]$ & $-0.24(3)$ & $-0.19(5)$ & $-0.24(5)$  & $-0.24(3)$ & $-0.19(5)$ & $-0.23(5)$\\
	$c_{\omega''} \ [\text{GeV}^{-1}]$ & $-1.69(5)$ & $-1.66(5)$ & $-1.58(5)$ & $-1.69(4)$ & $-1.66(5)$ & $-1.58(6)$\\
	$c_1 \ [\text{GeV}^{-3}]$ & $-0.25(6)$ & $-0.21(7)$ & $0.16(12)$ & $-0.25(5)$ & $-0.20(7)$ & $0.14(14)$\\
	$c_2 \ [\text{GeV}^{-3}]$ & $-1.21(3)$ & $-1.25(4)$ & $-1.51(7)$ & $-1.21(3)$ & $-1.25(4)$ & $-1.49(9)$\\
	$c_3 \ [\text{GeV}^{-3}]$ & --- & $-0.62(5)$ & $-0.57(5)$ & --- & $-0.61(5)$ & $-0.56(6)$\\
	$c_4 \ [\text{GeV}^{-3}]$ & --- & --- & $1.43(7)$ & --- & --- & $1.42(9)$ \\
	$10^4\times \xi_\text{CMD-2}$ & $1.3(5)$ & $1.3(5)$ & $1.3(5)$ & $1.3(5)$ & $1.3(5)$ & $1.3(5)$ \\
	$10^3\times \xi_\text{BaBar}$ & $1.3(2)$ & $1.3(1)$ & $1.3(1)$ & $1.3(1)$ & $1.3(2)$ & $1.3(2)$ \\
	$10^3\times \xi_\text{BaBar}' \ [\text{GeV}^{-1}]$ & $-2.3(3)$ & $-2.3(1)$ & $-2.3(1)$ & $-2.3(1)$ & $-2.3(3)$ & $-2.3(3)$ \\
	$10^3\times \Re\epsrw$ & $1.46(17)$ & $1.44(13)$ & $1.67(17)$ & $1.50(16)$ & $1.51(17)$ & $1.69(18)$\\
	$C_A/C_A^\text{WZW}$ &  $1.038(11)$ & $1.028(13)$ & $0.980(16)$ & $1.051(10)$ & $1.040(13)$ & $0.997(19)$\\
	$10^{10}\times a_\mu^{3\pi}|_{\leq 1.8\GeV}$ & $46.11(32)$ & $46.13(33)$ & $46.21(32)$ & $46.16(32)$ & $46.13(34)$ & $46.21(33)$\\
	$10^{10}\times a_\mu^\text{FSR}[3\pi]$ & $0.51(0)$ & $0.51(0)$ & $0.52(0)$ & $0.52(0)$ & $0.52(0)$ & $0.52(0)$\\
	$10^{10}\times a_\mu^{\rho\text{--}\omega}[3\pi]$ & $-2.64(30)$ & $-2.65(29)$ & $-3.02(29)$ & $-3.09(32)$ & $-3.10(36)$ & $-3.49(37)$\\
	\bottomrule
	\renewcommand{\arraystretch}{1.0}
	\end{tabular}
	\caption{Same as Table~\ref{tab:fits_global}, with the chiral anomaly as an additional fit parameter. $C_A^\text{WZW}$ is defined via Eqs.~\eqref{F3pi_WZW} and~\eqref{C_A_chiral}.}
	\label{tab:fits_global_CA}
\end{table}

\section{Extraction of chiral anomaly from global fit}
\label{sec:anomaly}

As reference point, we first reproduce the result of the global fit from Ref.~\cite{Hoferichter:2023bjm}, see Table~\ref{tab:fits_global}. While the quality of the combined fit is worse than that of fits to the individual data sets, suggesting some tensions in the data base, the required increase of the uncertainties by a scale factor $S=\sqrt{\chi^2/\text{dof}}\simeq 1.16$ remains rather moderate. To assess systematic uncertainties, we quoted the fits with number of free conformal parameters $p_\text{conf}=3$ for $\delta_\eps=0\degree$ as central value,\footnote{This choice is motivated by the fact that the $3\pi$ fits themselves are not sensitive enough to the phase, see Ref.~\cite{Hoferichter:2023bjm} for fits with a free $\delta_\eps$.} including the maximal variation to $p_\text{conf}=\{2,4\}$ and to $\delta_\eps=3.5\degree$ as additional uncertainty components besides the scale-factor-inflated statistical error.
For instance, for the contribution to the anomalous magnetic moment of the muon we obtained
\beq
a_\mu^{3\pi}|_{\leq 1.8\GeV}=45.91(37)(35)(13)[53]\times 10^{-10},
\eeq
with errors referring to statistics, order of conformal polynomial, and $\delta_\eps$, respectively, and the same error breakdown for other key quantities is reproduced in the first column of Table~\ref{tab:derived_quantities}.

One particularly important aspect of this global fit was that for the BaBar data~\cite{BABAR:2021cde}, taken using the initial-state-radiation (ISR) technique, an appropriate reweighting over the bins needed to be performed~\cite{Colangelo:2018mtw,Stamen:2022uqh,Hoferichter:2023bjm}, i.e.,
for a bin $\big[q^2_{i, \text{min}},q^2_{i, \text{max}}\big]$ the actual observable is given by
\beq
  f(x_i) = \frac{1}{q^2_{i, \text{max}}-q^2_{i, \text{min}}} \int_{q^2_{i, \text{min}}}^{q^2_{i, \text{max}}} \diff q^2\,  \sigma_{e^+e^-\to 3\pi(\gamma)}(q^2),
\eeq
which can also be implemented by
solving  $f(x_i)=\sigma_{e^+e^-\to3\pi(\gamma)}(q_i^2)$ for the correct bin centers. Without this correction, we were not able to find acceptable fits. In all cases, we implement an iterative fit strategy~\cite{Ball:2009qv} to avoid a D'Agostini bias~\cite{DAgostini:1993arp}.

\begin{table}[t]
	\centering
	\footnotesize
	\renewcommand{\arraystretch}{1.3}
	\begin{tabular}{lcccccc}
	\toprule
	& \multicolumn{3}{c}{$\delta_\eps=0\degree$} & \multicolumn{3}{c}{$\delta_\eps=3.5\degree$}\\
	$p_\text{conf}$ & $2$ & $3$ & $4$ & $2$ & $3$ & $4$\\
	$\chi^2/\text{dof}$ & $283.3/160$ & $281.2/159$ & $281.1/158$ & $283.8/160$ & $281.6/159$& $281.3/158$\\
	& $=1.77$ & $=1.77$ & $=1.78$ & $=1.77$ & $=1.77$& $=1.78$\\
	$p$-value & $7\times 10^{-9}$ & $8\times 10^{-9}$ & $6\times 10^{-9}$ & $6\times 10^{-9}$ & $7\times 10^{-9}$& $6\times 10^{-9}$\\
	$\mw \ [\text{MeV}]$ & $782.72(4)$ & $782.72(4)$ & $782.72(4)$ & $782.72(4)$ & $782.72(4)$& $782.72(4)$\\
	$\Gw \ [\text{MeV}]$ & $8.42(6)$ & $8.43(6)$ & $8.43(6)$  & $8.40(6)$ & $8.41(6)$& $8.42(6)$\\
	$\mphi \ [\text{MeV}]$& $1019.25(4)$ & $1019.24(4)$ & $1019.24(4)$ & $1019.25(4)$ & $1019.24(4)$& $1019.25(4)$\\
	$\Gphi \ [\text{MeV}]$ & $3.96(8)$ & $3.98(8)$ & $3.98(8)$  & $3.96(8)$ & $3.98(8)$& $3.98(8)$\\
	$M_{\omega'} \ [\text{GeV}]$ & $1.467(14)$ & $1.462(22)$& $1.459(28)$& $1.466(13)$& $1.461(21)$& $1.453(38)$\\
	$c_\omega \ [\text{GeV}^{-1}]$ & $2.99(4)$ & $2.99(4)$ & $2.99(4)$ & $2.99(5)$ & $2.99(5)$& $3.00(5)$\\
	$c_\phi \ [\text{GeV}^{-1}]$ & $-0.374(6)$ & $-0.376(6)$ & $-0.376(6)$ & $-0.374(6)$ & $-0.376(6)$& $-0.376(6)$\\
	$c_{\omega'} \ [\text{GeV}^{-1}]$ & $-0.34(6)$ & $-0.21(11)$ & $-0.21(9)$  & $-0.35(6)$ & $-0.22(11)$& $-0.21(11)$\\
	$c_{\omega''} \ [\text{GeV}^{-1}]$ & $-2.18(13)$ & $-2.04(16)$ & $-2.01(16)$ & $-2.22(13)$ & $-2.07(17)$& $-2.00(22)$\\
	$c_1 \ [\text{GeV}^{-3}]$ & $0.39(16)$ & $0.39(16)$ & $0.39(15)$ & $0.44(16)$ & $0.44(16)$& $0.44(17)$\\
	$c_2 \ [\text{GeV}^{-3}]$ & $-1.19(6)$ & $-1.28(8)$ & $-1.31(15)$ & $-1.18(6)$ & $-1.27(9)$& $-1.35(27)$\\
	$c_3 \ [\text{GeV}^{-3}]$ & --- & $-1.01(16)$ & $-0.98(17)$ & --- & $-1.05(17)$& $-0.97(30)$\\
	$c_4 \ [\text{GeV}^{-3}]$ & --- & --- & $1.54(11)$ & --- & --- & $1.56(14)$ \\
	$10^3\times \Re\epsrw$ & $1.57(43)$ & $1.50(44)$ & $1.52(44)$ & $1.58(44)$ & $1.51(45)$ & $1.56(47)$\\
	$10^{10}\times a_\mu^{3\pi}|_{\leq 1.8\GeV}$ & $48.52(1.04)$ & $48.65(1.05)$ & $48.67(1.02)$ & $48.41(1.03)$ & $48.54(1.04)$ & $48.59(1.06)$\\
	$10^{10}\times a_\mu^\text{FSR}[3\pi]$ & $0.54(1)$ & $0.54(1)$ & $0.54(1)$ & $0.54(1)$ & $0.54(1)$ & $0.55(1)$\\
	$10^{10}\times a_\mu^{\rho\text{--}\omega}[3\pi]$ & $-2.98(81)$ & $-2.87(82)$ & $-2.90(82)$ & $-3.42(95)$ & $-3.27(96)$ & $-3.39(1.02)$\\
	\bottomrule
	\renewcommand{\arraystretch}{1.0}
	\end{tabular}
	\caption{Fit to Belle II 2024~\cite{Belle-II:2024msd}, otherwise same as Table~\ref{tab:fits_global}.}
	\label{tab:fits_BelleII}
\end{table}

Normalizing $C_A$ to its WZW prediction, we obtain the results shown in Table~\ref{tab:fits_global_CA}. In particular, for the chiral anomaly we obtain
\beq
\label{CA_final}
\frac{C_A}{C_A^\text{WZW}}\bigg|_{\text{global fit}}=1.028(15)(48)(17)[53],
\eeq
in agreement with the expectation at a level of $5\%$. The uncertainty is dominated by the truncation of the conformal expansion, and indeed we observe that for larger $p_\text{conf}$, the fits become increasingly unstable, with marginal gains in the $\chi^2/\text{dof}$. One could, potentially, decrease the error by a more aggressive truncation, but we adhere to the same prescription formulated in Ref.~\cite{Hoferichter:2023bjm}. In this way, we obtain a central value for $C_A$ that lies $2.8\%$ above the chiral prediction, well within the estimated uncertainty of $5.3\%$, and thus providing a strong consistency check on the data base. Translating the result to the standard conventions for $F_{3\pi}$ via Eq.~\eqref{C_A_chiral}, we obtain
\beq
F_{3\pi}^\text{global fit}=33.1(1.7)\GeV^{-3},
\eeq
where the uncertainty in the quark-mass correction is still negligible compared to the dominant truncation error.

\begin{table}[t]
	\centering
	\footnotesize
	\renewcommand{\arraystretch}{1.3}
	\begin{tabular}{lcccccc}
	\toprule
	& \multicolumn{3}{c}{$\delta_\eps=0\degree$} & \multicolumn{3}{c}{$\delta_\eps=3.5\degree$}\\
	$p_\text{conf}$ & $2$ & $3$ & $4$ & $2$ & $3$ & $4$\\
	$\chi^2/\text{dof}$ & $276.6/160$ & $274.7/159$ & $274.6/158$ & $278.5/160$ & $276.5/159$& $276.2/158$\\
	& $=1.73$ & $=1.73$ & $=1.74$ & $=1.74$ & $=1.74$& $=1.75$\\
	$p$-value & $3\times 10^{-8}$ & $3\times 10^{-8}$ & $3\times 10^{-8}$ & $2\times 10^{-8}$ & $2\times 10^{-8}$& $2\times 10^{-8}$\\
	$\mw \ [\text{MeV}]$ & $782.76(4)$ & $782.75(4)$ & $782.75(4)$ & $782.76(4)$ & $782.75(4)$& $782.75(4)$\\
	$\Gw \ [\text{MeV}]$ & $8.55(7)$ & $8.56(6)$ & $8.56(6)$  & $8.53(6)$ & $8.53(6)$& $8.54(6)$\\
	$\mphi \ [\text{MeV}]$& $1019.40(4)$ & $1019.39(4)$ & $1019.39(4)$ & $1019.40(4)$ & $1019.39(4)$& $1019.40(4)$\\
	$\Gphi \ [\text{MeV}]$ & $4.42(8)$ & $4.44(8)$ & $4.43(8)$  & $4.42(8)$ & $4.44(8)$& $4.43(8)$\\
	$M_{\omega'} \ [\text{GeV}]$ & $1.471(13)$ & $1.467(19)$& $1.465(30)$& $1.469(13)$& $1.466(18)$& $1.460(30)$\\
	$c_\omega \ [\text{GeV}^{-1}]$ & $3.04(4)$ & $3.04(4)$ & $3.04(5)$ & $3.04(4)$ & $3.04(5)$& $3.05(5)$\\
	$c_\phi \ [\text{GeV}^{-1}]$ & $-0.396(6)$ & $-0.397(6)$ & $-0.397(6)$ & $-0.395(6)$ & $-0.397(6)$& $-0.397(6)$\\
	$c_{\omega'} \ [\text{GeV}^{-1}]$ & $-0.34(6)$ & $-0.22(11)$ & $-0.21(12)$  & $-0.35(6)$ & $-0.23(11)$& $-0.21(12)$\\
	$c_{\omega''} \ [\text{GeV}^{-1}]$ & $-2.18(13)$ & $-2.05(16)$ & $-2.02(9)$ & $-2.23(12)$ & $-2.09(17)$& $-2.02(27)$\\
	$c_1 \ [\text{GeV}^{-3}]$ & $0.39(16)$ & $0.39(15)$ & $0.38(16)$ & $0.45(15)$ & $0.45(15)$& $0.44(16)$\\
	$c_2 \ [\text{GeV}^{-3}]$ & $-1.20(6)$ & $-1.28(9)$ & $-1.31(27)$ & $-1.18(6)$ & $-1.27(9)$& $-1.36(25)$\\
	$c_3 \ [\text{GeV}^{-3}]$ & --- & $-1.07(16)$ & $-1.03(30)$ & --- & $-1.11(17)$& $-1.03(28)$\\
	$c_4 \ [\text{GeV}^{-3}]$ & --- & --- & $1.56(14)$ & --- & --- & $1.59(14)$ \\
	$10^3\times \Re\epsrw$ & $2.14(43)$ & $2.07(43)$ & $2.09(44)$ & $2.12(44)$ & $2.04(44)$ & $2.09(46)$\\
	$10^{10}\times a_\mu^{3\pi}|_{\leq 1.8\GeV}$ & $48.40(1.04)$ & $48.50(1.04)$ & $48.53(1.06)$ & $48.24(1.03)$ & $48.35(1.03)$ & $48.40(1.05)$\\
	$10^{10}\times a_\mu^\text{FSR}[3\pi]$ & $0.55(1)$ & $0.55(1)$ & $0.55(1)$ & $0.55(1)$ & $0.55(1)$ & $0.56(1)$\\
	$10^{10}\times a_\mu^{\rho\text{--}\omega}[3\pi]$ & $-4.01(77)$ & $-3.90(80)$ & $-3.93(80)$ & $-4.54(92)$ & $-4.39(94)$ & $-4.51(98)$\\
	\bottomrule
	\renewcommand{\arraystretch}{1.0}
	\end{tabular}
	\caption{Same as Table~\ref{tab:fits_BelleII}, without reweighting over the bins.}
	\label{tab:fits_BelleII_no_reweighting}
\end{table}

\section{Fits to Belle II}
\label{sec:BelleII}

Belle II released data for $e^+e^-\to 3\pi$ in Ref.~\cite{Belle-II:2024msd}, thereof $174$ data points below $1.8\GeV$, including statistical and systematic covariance matrices. We performed fits of our dispersive representation to a subset of $172$ data points, removing two points with negative central values for the cross section.\footnote{Due to the large errors, the impact of these points on the fit is minimal. In the vector-meson-dominance (VMD) fits performed in Ref.~\cite{Belle-II:2024msd}, instead a cut on the singular values of the inverse covariance matrix is imposed~\cite{Sue}. After removing these two points, a cut with the same threshold no longer becomes active, strongly suggesting that both approaches are equivalent.} The results of this fit are given in Table~\ref{tab:fits_BelleII}.

\begin{table}[t]
	\centering
	\footnotesize
	\renewcommand{\arraystretch}{1.3}
	\begin{tabular}{lcccccc}
	\toprule
	& \multicolumn{3}{c}{$\delta_\eps=0\degree$} & \multicolumn{3}{c}{$\delta_\eps=3.5\degree$}\\
	$p_\text{conf}$ & $2$ & $3$ & $4$ & $2$ & $3$ & $4$\\
	$\chi^2/\text{dof}$ & $278.5/159$ & $277.2/158$ & $276.2/157$ & $277.3/159$ & $276.1/158$& $275.1/157$\\
	& $=1.75$ & $=1.75$ & $=1.76$ & $=1.74$ & $=1.75$& $=1.75$\\
	$p$-value & $1\times 10^{-8}$ & $1\times 10^{-8}$ & $1\times 10^{-8}$ & $2\times 10^{-8}$ & $2\times 10^{-8}$& $2\times 10^{-8}$\\
	$\mw \ [\text{MeV}]$ & $782.71(4)$ & $782.71(4)$ & $782.71(4)$ & $782.72(4)$ & $782.72(4)$& $782.71(4)$\\
	$\Gw \ [\text{MeV}]$ & $8.49(7)$ & $8.49(7)$ & $8.50(7)$  & $8.49(7)$ & $8.49(7)$& $8.50(7)$\\
	$\mphi \ [\text{MeV}]$& $1019.25(4)$ & $1019.24(4)$ & $1019.24(4)$ & $1019.25(4)$ & $1019.25(4)$& $1019.24(4)$\\
	$\Gphi \ [\text{MeV}]$ & $3.96(8)$ & $3.97(8)$ & $3.97(8)$  & $3.96(8)$ & $3.97(8)$& $3.98(8)$\\
	$M_{\omega'} \ [\text{GeV}]$ & $1.468(20)$ & $1.463(28)$& $1.490(22)$& $1.467(20)$& $1.462(28)$& $1.489(22)$\\
	$c_\omega \ [\text{GeV}^{-1}]$ & $3.00(5)$ & $3.01(5)$ & $3.00(5)$ & $3.02(5)$ & $3.03(5)$& $3.02(5)$\\
	$c_\phi \ [\text{GeV}^{-1}]$ & $-0.376(6)$ & $-0.377(6)$ & $-0.378(6)$ & $-0.376(6)$ & $-0.377(6)$& $-0.379(6)$\\
	$c_{\omega'} \ [\text{GeV}^{-1}]$ & $-0.24(10)$ & $-0.17(11)$ & $-0.26(15)$  & $-0.24(10)$ & $-0.18(11)$& $-0.26(15)$\\
	$c_{\omega''} \ [\text{GeV}^{-1}]$ & $-1.79(27)$ & $-1.79(24)$ & $-2.08(35)$ & $-1.80(27)$ & $-1.80(24)$& $-2.08(35)$\\
	$c_1 \ [\text{GeV}^{-3}]$ & $-0.12(36)$ & $0.03(32)$ & $-0.02(27)$ & $-0.11(35)$ & $0.03(32)$& $-0.01(27)$\\
	$c_2 \ [\text{GeV}^{-3}]$ & $-1.23(6)$ & $-1.28(8)$ & $-0.93(37)$ & $-1.23(6)$ & $-1.28(8)$& $-0.93(37)$\\
	$c_3 \ [\text{GeV}^{-3}]$ & --- & $-0.71(27)$ & $-0.99(38)$ & --- & $-0.71(27)$& $-0.99(37)$\\
	$c_4 \ [\text{GeV}^{-3}]$ & --- & --- & $1.20(23)$ & --- & --- & $1.20(23)$ \\
	$10^3\times \Re\epsrw$ & $1.38(44)$ & $1.40(43)$ & $1.21(47)$ & $1.47(44)$ & $1.49(43)$ & $1.30(46)$\\
	$C_A/C_A^\text{WZW}$ &  $1.087(46)$ & $1.069(43)$ & $1.098(50)$ & $1.098(45)$ & $1.081(43)$ & $1.108(48)$\\
	$10^{10}\times a_\mu^{3\pi}|_{\leq 1.8\GeV}$ & $49.01(1.10)$ & $49.07(1.10)$ & $49.10(1.10)$ & $49.02(1.10)$ & $49.08(1.10)$ & $49.10(1.10)$\\
	$10^{10}\times a_\mu^\text{FSR}[3\pi]$ & $0.54(1)$ & $0.54(1)$ & $0.54(1)$ & $0.55(1)$ & $0.55(1)$ & $0.55(1)$\\
	$10^{10}\times a_\mu^{\rho\text{--}\omega}[3\pi]$ & $-2.66(84)$ & $-2.71(82)$ & $-2.34(90)$ & $-3.23(96)$ & $-3.27(95)$ & $-2.86(1.02)$\\
	\bottomrule
	\renewcommand{\arraystretch}{1.0}
	\end{tabular}
	\caption{Same as Table~\ref{tab:fits_BelleII}, with the chiral anomaly as an additional fit parameter.}
	\label{tab:fits_BelleII_CA}
\end{table}

The main observations are the following: (i) The fit quality is significantly worse than for the other data sets, in fact, much worse than even for the global fit (as described in Ref.~\cite{Hoferichter:2023bjm}, the $p$-values of the fits to the other individual data sets are at least a few permil). This clearly points to tensions with the dispersive constraints incorporated into our representation. (ii) While $\mw$, $\mphi$, $M_{\omega'}$, and $\Re\epsrw$ come out consistent, the deficit in $\Gw$ and $\Gphi$ evaluates to $3.2\sigma$ and $2.6\sigma$, respectively. The tension in $a_\mu^{3\pi}$ amounts to $1.8\sigma$, compared to the $2.5\sigma$ tension quoted in Ref.~\cite{Belle-II:2024msd} based on $a_\mu^{3\pi}= 48.91(1.09)\times 10^{-10}$. This difference is mainly driven by the scale factor $S=\sqrt{1.77}=1.33$ as well as the slightly lower central value in our result.

\begin{figure}[t]
	\centering
	\includegraphics[width=0.7\linewidth]{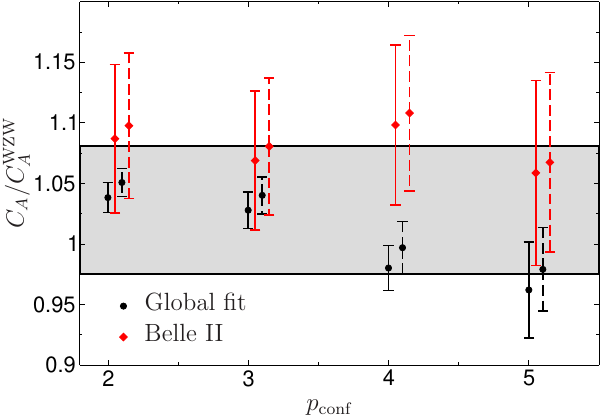}
	\caption{Extracted value of $C_A/C_A^\text{WZW}$ as a function of $p_\text{conf}$ for the global fit (circles, black) and Belle II (diamonds, red), with error bars reflecting the statistical uncertainties including scale factors. The solid points refer to $\delta_\eps=0\degree$, the dashed ones to $\delta_\eps=3.5\degree$. The gray band indicates our final result~\eqref{CA_final}.
	}
	\label{fig:CA}
\end{figure}

This distortion of the $\omega$ and $\phi$ line shapes, with widths significantly smaller than the literature values,\footnote{Note that in comparison to the conventions from Ref.~\cite{ParticleDataGroup:2024cfk}, VP corrections need to be applied, i.e., to obtain the vector-meson parameters including VP from the ones given here, the shifts $\Delta \mw=0.13\MeV$, $\Delta \mphi=0.26\MeV$, $\Delta \Gw=-0.06\MeV$, $\Delta \Gphi\simeq 0$~\cite{Hoferichter:2019mqg,Holz:2022hwz} need to be added. To obtain the bare cross sections, we used the VP routine from Ref.~\cite{Keshavarzi:2018mgv}, but checked that the difference to the VP routine~\cite{Ignatov:VP}, as employed in Ref.~\cite{Belle-II:2024msd}, only leads to negligible changes (see also Ref.~\cite{Aliberti:2024fpq} for the comparison).} was not observed in the VMD fit of Ref.~\cite{Belle-II:2024msd} (without reweighting), which is why we repeated our fits without the reweighting over the bins as well, see Table~\ref{tab:fits_BelleII_no_reweighting}. For both cases, the key quantities including their error breakdown are again summarized in Table~\ref{tab:derived_quantities}. We observe that indeed the tensions in $\Gw$ and $\Gphi$ are largely mitigated, at the expense of introducing a tension in $\mphi$ at the level of $3.5\sigma$. These findings agree with the VMD fits from Ref.~\cite{Belle-II:2024msd}, suggesting that the difference is driven by the reweighting over the bin. In contrast to the BaBar fits, in this case the fit quality with and without reweighting is almost identical, although conceptually it is clear that especially in the vicinity of the sharp resonance peaks the distribution of events within a given bin needs to be taken into consideration. We therefore consider both the tensions with the dispersive representation and in the vector-meson parameters as issues in the interpretation of the data set that should be taken seriously.

Finally, we also repeated the fits with $C_A$ as a free parameter for the Belle II data, see Table~\ref{tab:fits_BelleII_CA} for the results. In the same way as for the global fit, the fit quality improved marginally, but remains at a level that is not acceptable. Moreover, the tensions in $\Gw$, $\Gphi$ persist, while the one in $a_\mu^{3\pi}$ even increases a little. The result for the chiral anomaly becomes
\beq
\frac{C_A}{C_A^\text{WZW}}\bigg|_\text{Belle II}=1.069(58)(29)(12)[66],
\eeq
consistent with the expectation within uncertainties, but exhibiting a clear upwards pull especially when considering that the results for $p_\text{conf}=\{2,4\}$ are larger in all cases, see Table~\ref{tab:fits_BelleII_CA}. This behavior is consistent with the tension in $a_\mu^{3\pi}$, essentially reflecting the same upwards pull in the cross section.

The results for $C_A/C_A^\text{WZW}$ are illustrated in Fig.~\ref{fig:CA}, up to order $p_\text{conf}=5$. One can see again that the results become less stable for large degrees of the conformal polynomial, as indicated by the increasing statistical errors and marginal gains in the $\chi^2/\text{dof}$. Moreover, one can see that the statistical errors of Belle II are significantly larger than for the global fit, partly because of the scale-factor inflation. While consistent within uncertainties, the figure illustrates how the upward tension in $a_\mu^{3\pi}$ is mirrored by an upward pull in the chiral anomaly as well.

\section{Conclusions}
\label{sec:conclusions}

Based on a dispersive representation for the $e^+e^-\to 3\pi$ cross section, we presented an extraction of the chiral anomaly $F_{3\pi}$ from cross-section data, leveraging the fact that via  a sum rule the chiral anomaly is tied towards the $\omega$ and $\phi$ resonances, thus significantly enhancing the sensitivity beyond what would otherwise be possible in the low-energy region alone. Studying the sensitivity to the systematic uncertainties in our representation, most notably the truncation of a conformal polynomial and the phase of the $\rho$--$\omega$ mixing parameter, we found that $F_{3\pi}$ can be extracted with a precision of about $5\%$, with a central value well compatible with the anomaly prediction, see Eq.~\eqref{CA_final} and Fig.~\ref{fig:CA}. These findings provide a welcome consistency check on the $e^+e^-\to 3\pi$ data base and support the standard assumption to fix the subtraction term via the chiral anomaly. Our result, $F_{3\pi}=33.1(1.7)\GeV^{-3}$ also agrees well with the preliminary result from COMPASS, $F_{3\pi}= 34.0(2.0)\GeV^{-3}$~\cite{Ecker:2023qae}, and defines a useful point of comparison for their final analysis.

We also considered fits to the recent $e^+e^-\to3\pi$ data from Belle II, in this case finding tensions with the dispersive constraints, the width parameters of $\omega$ and $\phi$, and the resulting $3\pi$ contribution to the anomalous magnetic moment of the muon, the latter being reflected by an upward pull in the chiral anomaly. While the statistical errors at this stage are still fairly sizable and thus the significance not dramatic, we believe these tensions should be taken seriously and closely monitored once more data become available.

\acknowledgments
We thank  Andrei Rabusov for suggesting to try and extract the chiral anomaly from $e^+e^-\to 3\pi$, as well as
Yuki Sue and James Libby for detailed discussions and correspondence on Ref.~\cite{Belle-II:2024msd}.
Financial support by the SNSF (Project No.\ TMCG-2\_213690) and the DFG through the fund provided to the Research Unit  ``Photon--photon interactions in the Standard Model and beyond'' (Projektnummer 458854507 -- FOR 5327) is gratefully acknowledged.


\bibliographystyle{apsrev4-1_mod_2}
\bibliography{amu}

\end{document}